# A Rprop-Neural-Network-Based PV Maximum Power Point Tracking Algorithm with Short-Circuit Current Limitation

Yao Cui[1,2], Zhehan Yi[1], Jiajun Duan[1], Di Shi[1], Zhiwei Wang[1]
[1]GEIRI North America, San Jose, CA 95134
[2]The George Washington University, Washington, DC 20052
Emails: raven_cuiyao@gwu.edu, {zhehan.yi, jiajun.duan, di.shi, zhiwei.wang}@geirina.net

*Abstract*—This paper proposes a resilient-backpropagation-neural-network-(Rprop-NN) based algorithm for Photovoltaic (PV) maximum power point tracking (MPPT). A supervision mechanism is proposed to calibrate the Rprop-NN-MPPT reference and limit short-circuit current caused by incorrect prediction. Conventional MPPT algorithms (e.g., perturb and observe (P&O), hill climbing, and incremental conductance (Inc-Cond) etc.) are trial-and-error-based, which may result in steady-state oscillations and loss of tracking direction under fast changing ambient environment. In addition, partial shading is also a challenge due to the difficulty of finding the global maximum power point on a multi-peak characteristic curve. As an attempt to address the aforementioned issues, a novel Rprop-NN MPPT algorithm is developed and elaborated in this work. Multiple case studies are carried out to verify the effectiveness of proposed algorithm.

*Index Terms*— *Rprop Neural network, maximum power point tracking, short-circuit current, partial shading, PV.*

## I. Introduction

Along with the significant increasing of global energy demand, the need of renewable energy is incredibly raised due to the considerations of environmental impact and sustainable development [1]. At the same time, PV generation have been rapidly increased for the past two decades and become one of the most important clean resources because of its desirable characteristics such as high availability, scalability, energy independency, and low maintenance cost [2-5]. These remarkable advantages make it even more promising for large scale utility application.

However, the maximum power that a PV array can generate depends highly on the operational environment, e.g., temperature and solar irradiance. To improve the utilization of solar energy, MPPT algorithms, which are typically implemented in the controllers of PV, are designed to extract the maximum power of a PV array under various conditions. In the literature, quite a few methods have been proposed to realize the PV MPPT through advanced power-electronics as well as software techniques [6]. In practice, P&O and Inc-Cond based algorithms are the most widely employed [7, 8], due to their mechanism and implementation simplicity [9-12]. However, these methods rely on a trial-and-error-based maximum finding mechanism that leads to the following major drawbacks:

1) Steady-state oscillation: trial-and-error-based methods cannot converge to the exact optimal operating point, and keep going forth and back and continuously oscillating around the control set point [13-16];

2) Loss of tracking direction: under fast irradiance fluctuation conditions, the PV operational point trajectory may diverge from the actual maximum power point gradually [17, 18].

As the penetration level of PV power increases in the bulk power system, these issues may introduce significant disturbances to the grid. Therefore, as an attempt to address these challenges, a RPROP-NN-based MPPT method with supervision short-circuit current limitation capability is proposed in this paper. The proposed method predicts the maximum power point of a PV array instantaneously based on the real-time irradiance and temperature measurements, which allows the PV system to move to the optimal operational point directly without trial-and-error behaviors and eliminates the steady oscillations and loss of tracking direction. An adaptive learning rate is designed to guarantee a fast converge speed of the proposed algorithm. To avoid over-prediction, for instance, under partial shading conditions, a supervision mechanism is devised to eliminate the false short circuit scenario and lead the PV system to converge on the actual optimal operational point.

The rest of this paper is organized as follows: Section II analyzes the detailed drawback of conventional trial-and-error-based MPPT methods; Section III elaborates the proposed RPROP-NN based MPPT algorithm; Multiple case studies are presented in Section IV to evaluate the proposed method; and Section V concludes this paper.

## II. Challneges for Conventional MPPT

This section aims at elaborating the issues for conventional trial-and-error-based PV MPPT methods.

### A. Steady-State Oscillation

The power-voltage (P-V) characteristic curve of a PV array is shown in Fig. 1 (a), where the MPP stands for maximum power point. Most of the conventional MPPT methods are trial-and-error-based, which might be modified from P&O and Inc-Cond algorithms. These methods will introduce a steady-state oscillation when the PV system reaches the MPP. For the sake of a clear presentation, P&O method is used to explain the

This work is supported by the SGCC Science and Technology Program *Distributed High-Speed Frequency Control under UHVDC Bipolar Blocking Fault Scenario*.

issue. The basic idea of P&O algorithm is to find the optimal operation point of a PV array on the via a hill-climbing-based approach. For instance, the array is firstly working at point $a$ with the voltage and power measurements, $V_a$ and $P_a$, respectively. The controller perturbs the operating point from $a$ to $b$ and measurements the voltage and power $V_b$ and $P_b$. If $P_b > P_a$ and $V_b > V_a$, the controller will move the operating point from $a$ to $b$ and repeat the same procedures to find the MPP. However, when it reaches MPP, the controller keeps perturbing the operating points to detect system changes, e.g., irradiance change, which leads to constant power and voltage oscillations [13].

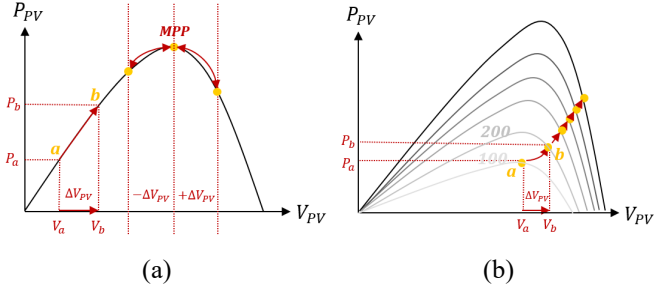

(a)               (b)

Fig. 1 Illustration of conventional MPPT issues: (a) steady-state oscillation, (b) loss of tracking directions.

### B. Loss of Tracking Direction

Another major problem of conventional trial-and-error-based MPPT algorithms is the loss of tracking direction, especially under fast irradiance changing situations. Fig. 1 (b) illustrates this process, where multiple P-V curves are plotted for different irradiance levels. For instance, when the MPPT controller perturbs the operating voltage from $V_a$ to $V_b$ while the solar irradiance increases from 100 W/m² to 200 W/m², the controller measures the powers and find $P_b > P_a$. It will move the operating point from $a$ to $b$. If the irradiance keeps increasing in this process, the controller will follow the same procedures and lead the operating point to diverge from the actual MPP. In a cloudy situation, the solar irradiance may fluctuate rapidly, which will lead to MPPT underperformance and hinder the full potentials of the PV array.

## III. THE PROPOSED RPROP-NN MPPT FRAMEWORK

### A. System Modeling

Fig. 2 illustrates a typical configuration of a grid-connected PV system [19] which consists of a PV array, a DC/DC converter, a DC/AC voltage source converter (VSC), and a transformer. The Rprop-NN MPPT algorithm is implemented in the controller of the DC/DC converter [20]. Real-time measurements of operating environment (i.e. irradiance and temperature) and PV parameters are fed into the Rprop-NN, which aggregates the data and predicts the optimal operating points (MPP reference) for the present condition (Fig. 2).

### B. Rprop-NN for PV MPPT

The proposed Rprop-NN MPPT method consists of a four-layer NN including one input layer, one output layer, and two hidden layers with 20 neurons in each layer. The back-propagation algorithm is used to calculate the gradient that is needed for updating the weights of NN. Through numerical offline simulations, multiple P-V characteristic curves of the tested PV array at different irradiance and temperature levels can be collected. Then, the vertices values can be mathematically found, which corresponds to the MPPs at different operating environments. Then, these data sets are applied as training samples to the NN in which two inputs (i.e., irradiance and temperature) and one output (i.e., MPP reference) are considered.

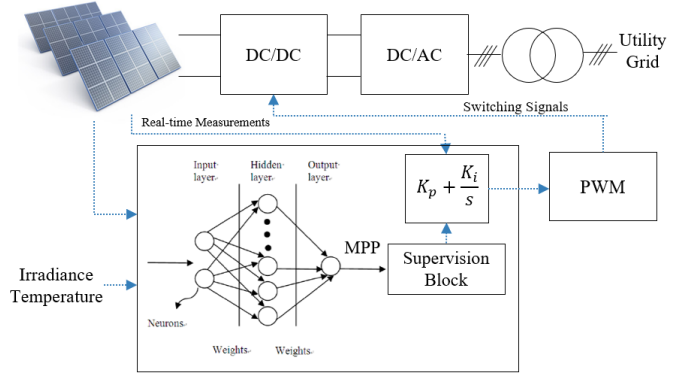

Fig. 2 Framework of the proposed method.

The '*trainrp*' function has been applied to train the network and to find the minimum point of the loss function with an adaptive learning rate.

The calculation process of this neural network could be divided into two parts: forward propagation processing and back propagation processing. During the forward propagation processing, the NN propagates layer by layer using (1) to get the output y,

$$\begin{cases} z^l = \omega^l \cdot f_l(z^{l-1}) + b^l \\ y = f_l(z^l) \end{cases} \quad (1)$$

where $z^l$, $\omega^l$, $f_l(\bullet)$ and $b^l$ are the inputs, weight matrix, activation function for neurons of the layer $l$, $b^l$ represents bias from layer $l-1$ to layer $l$, respectively. Accordingly, the square error $E$ can be calculated by

$$E = \Sigma \frac{1}{2}(y^* - y)^2 \quad (2)$$

where $y^*$ is the expected output.

Then, the NN enters the back propagation process by updating weights $\omega$ and bias $b$ from output layer to input layer following the direction of gradient descent. The updating laws can be defined as

$$\omega_{t+1} = \omega_t - \mu \frac{\partial E}{\partial \omega_t} \quad (3)$$

$$b_{t+1} = b_t - \mu \frac{\partial E}{\partial b_t} \quad (4)$$

where $\mu$ is the learning rate of weight, $\omega_{t+1}$ and $b_{t+1}$ are updatedweight and bias, respectively.

It is noteworthy to mention that an improvement has been made for the proposed Rprop-NN comparing to the conventional method. Instead of determining the changing amount of the weight, the partial derivative of loss function determines the changing direction on weights. If the signs of gradients of the loss function are different at the time point *t-1* and *t*, which means it has crossed the minimum point of the function, the learning rate $\mu$ should multiply a constant $\mu^{down}$ ($0 < \mu^{down} < 1$). On the contrary, if the symbols are the same, namely it has not reached the minimum point the, the $\mu$ can multiply a constant $\mu^{up}$ ($\mu^{up} > z$). Comparing to the traditional algorithms, it leads to a faster convergence speed, and there is no need to set parameters in advance so that avoid the difficulty of setting an exact optimal learning rate.

The corresponding updating algorithm can be mathematically presented as

$$\mu(t) = \begin{cases} \mu^{up}\mu_{t-1}, \forall g(t-1)g(t) > 0 \\ \mu^{down}\mu_{t-1}, \forall g(t-1)g(t) < 0 \\ \mu, otherwise \end{cases} \quad (5)$$

where $g(t)$ is the gradient of the error function at time *t*. The detailed implementation process of the proposed Rprop-NN method is illustrated in Table I.

TABLE I  IMPLEMENTATION OF THE NEURAL NETWORK

1. Normalize input value $G$, $T$
2. **for** neuron *1* to *j* from layer *l*,
   calculate the output using (1)
   calculate the square error E using (2)
3. **if** $E > \varepsilon$
   update weight $\omega$ and bias *b* using adaptive learning rate based on (3) to (5)
4. **else** output $f_l(z^l)$
5. **end**

Whenever irradiance or temperature changes, the pre-trained Rprop-NN MPPT module will collect data and predict a corresponding maximum power reference. Next, a classic proportional-integral (PI) method is utilized to control the DC/DC converter to make PV arrays tracking the updated maximum power reference. The external grid is responsible for regulating the DC-link voltage through a VSC.

The proposed Rprop-NN based method is able to predict and track the MPP in real time regardless of significant changes in irradiance and temperature, which avoids the common drawbacks of conventional MPPT methods. However, in practical applications, the training data sets collected from the simulation model can be different from the actual system due to panel aging. Then, the PV system may not be able to achieve the maximum power predicted by the Rprop-NN based on existing information. Furthermore, partial shading is another important challenge for solar energy. If solar panel is partially shaded, the power-current curve will appear multiple peaks and it becomes quite challenging to distinguish the global maximum and local maximum. To address this, a supervision algorithm is proposed in next subsection to tune the MPP reference in real time.

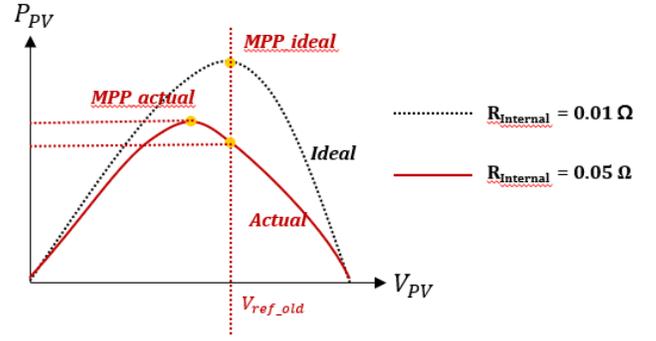

Fig. 3  Illustration of impact of internal impedance on MPPT.

*C. Supervised Short-Circuit Limitation*

Fig. 3 illustrates the phenomenon of above-mentioned problems. As can be seen, when PV aging issue occurs, the increased internal series resistance value decreases the maximum power that it can generate. Accordingly, the reference given by the Rprop-NN is greater than the actual power. As a result, a "pseudo short circuit" phenomenon will occur when system cannot reach and track the given reference, and the output current of the PV array will rush to its short-circuit current. Therefore, the output current of PV array can be considered as an event trigger to design the output current supervision algorithm:

In general, the real-time short-circuit current can be calculated based on the information of irradiance and temperature

$$Isc_{real} = Isc_{stc} \cdot \left(\frac{G}{1000}\right)^{1.01} \cdot \left(\frac{T(K)}{300}\right)^{0.2775} \quad (6)$$

where the $Isc_{stc}$ is the short-circuit current under standard test conditions. $G$ is irradiance and $T(K)$ represents temperature. When the $Isc_{real}$ is larger than 94% of short-circuit current, the Rprop-NN output y will multiply an attenuation coefficient $\gamma$ to get the calibrated reference $y_{cor}$

$$y_{cor} = \gamma y \quad (7)$$

Once the actual maximum power is reached, the current will return to the normal value. In other words, the proposed method can lead the system to find the actual MPP from top to bottom. Note that conventional trial-and-error based MPPT algorithms explore the MPP from bottom to top, which will result in a local instead of global MPP. From this perspective, the proposed supervised Rprop-NN method is able to locate the exact global MPP, since it explores the feasible maximum points from top to bottom. The detailed implementation process of the supervision block is illustrated in Table II.

TABLE II. IMPLEMENTATION OF THE SUPERVISION BLOCK

1. Calculate the $Isc_{real}$ using (6)
2. **if** $I > 0.94 \cdot Isc_{real}$
   update $y_{cor}$ based on (7)
3. **else** output $y_{cor}$
4. **end**

## IV. CASE STUDIES

In this section, multiple case studies are carried out based on the proposed Rprop-NN MPPT model with the supervised short-circuit limitation algorithm. The simulation is carried out in a grid-connected PV system as shown in Fig. 2 and Table III lists the parameters and values.

TABLE III. CASE STUDY SYSTEM PARAMETERS

| Parameter | Value | Description |
| --- | --- | --- |
| $\mu^{up}$ | 1.2 | Growth coefficient |
| $\mu^{down}$ | 0.5 | Attenuation coefficient |
| $Isc_{stc}$ | 1500 A | Short-circuit current under standard test conditions |
| $\gamma_{case\ I}$ | 0.95 | Attenuation coefficient |
| $\gamma_{case\ II}$ | 0.479 | Attenuation coefficient |
| $V_{oc}$ | 10 V | Open-circuit voltage |
| $I_{sc}$ | 15 A | Short-circuit current |
| $V_{mpp}$ | 8.25 V | Voltage at maximum power point |
| $P_{max}$ | 115.5 W | Maximum power point value |

### A. Case I. MPPT Performance to the Change of Irradiance and Temperature

The irradiance and temperature curves are plotted in Fig. 4, which are used as the inputs for both PV array and Rprop-NN MPPT controller. The performance of the proposed MPPT method is shown in Fig. 5, where the blue and red lines represents the actual and reference powers, respectively. It can be observed that the proposed Rprop-NN finds the MPP in real time, which is tracked by the actual power. Furthermore, between the time 1s and 2s, the reference value manifests oscillations to adapt a rapid change in the irradiance level as shown in Fig. 4. Nevertheless, the actual power still changes smoothly, which verifies the effectiveness of this method.

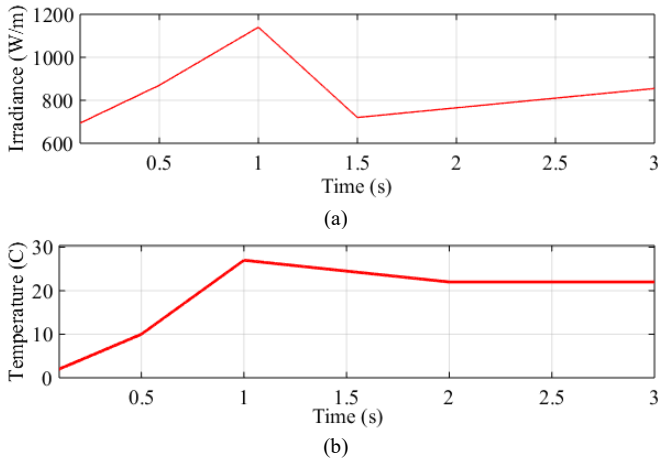

Fig. 4 Changing pattern of ambient environment: (a) real-time irradiance, (b) real-time temperature.

### B. Case II. Partial Shading Problem

To verify the performance of the proposed method in partial shading condition, half of the PV array is given the irradiance of 1000 W/m², while the other half is given 500 W/m² so as to simulate the shade. The performance of the proposed MPPT method is shown in Fig. 6, where the blue and red curves represent the actual and reference powers, respectively. Under partial shading condition, the predicted MPP from Rprop-NN is adjusted automatically by the supervision block to make sure the amount of power can be generated. In steady state, the actual power converges smoothly under oscillating references. Characteristic curves with multiple peaks for partially-shaded PV system is presented in Fig 7. It should be noted that the actual power generation is slightly less than the theoretical MPP because of system losses.

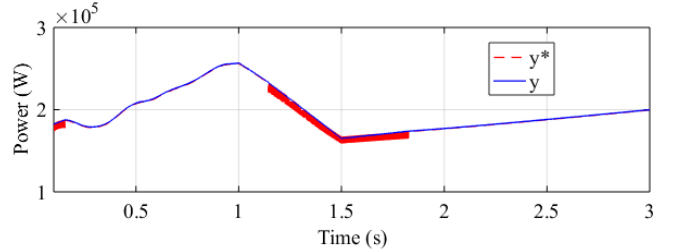

Fig. 5 Tracking performance of actual output power along with Rprop-NN MPPT reference.

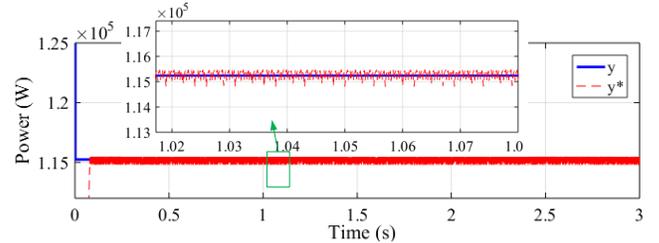

Fig. 6 Tracking performance of actual output power along with reference uder the partial shading condition.

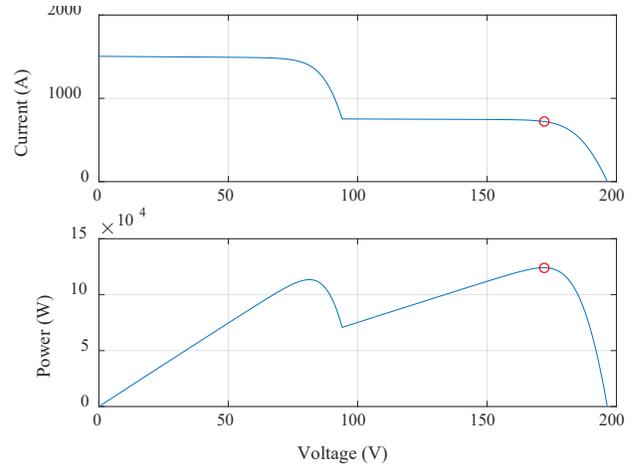

Fig. 7 Characteristic curves of the partially shaded PV system.

### C. Case III. Benchmarking Comparison with Conventional MPPT

To verify the performance of the proposed method, benchmarking comparisons are carried out by simulating the same PV system in the PSCAD/EMTDC software packages using the proposed method and traditional Inc-Cond algorithm. Two cases are tested: irradiance change from 300 to 1000 W/m² in 10 seconds and 2 seconds, respectively. The simulation results

(power-voltage characteristic curves) are shown in Fig. 8 and Fig. 9. As is shown the figures, conventional Inc-Cond MPPT suffers from significant power and voltage oscillations that lead to a continuous fluctuating operating point movement (Fig. 8). The red curves present the trajectory of the operating point. However, the proposed Rprop-NN method presents a clear trajectory with a stable and fast MPP tracking (Fig. 9).

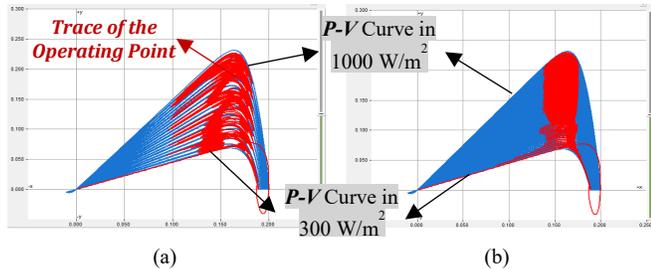

(a)          (b)

Fig. 8 Operating point trajectory using Inc-Cond MPPT: (a) irradiance changes from 300 to 1000 W/m$^2$ in 2 seconds. (b) irradiance: irradiance changes from 300 to 1000 W/m$^2$ in 10 seconds.

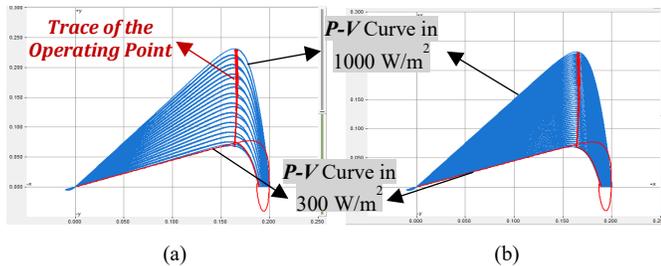

(a)          (b)

Fig. 9 Operating point trajectory using the Rprop-NN MPPT: (a) irradiance changes from 300 to 1000 W/m$^2$ in 2 seconds. (b) irradiance changes from 300 to 1000 W/m$^2$ in 10 seconds.

## V. CONCLUSION

This paper proposes a novel MPPT method based on Rprop-NN with supervised short-circuit current limitation. The proposed method predicts the real-time MPP of a PV array using measurements of irradiance and temperature. The paper firstly analyzes the drawbacks and causes of the conventional MPPT algorithms and devises the Rprop-NN MPPT that eliminates steady-state oscillations and loss of tracking direction issues, which also effectively improves the MPPT accuracy for aged and partially shaded PV arrays. The proposed algorithm has been demonstrated in simulations and performance evaluations have been carried out with benchmarking comparisons. Moreover, multiple case studies have verified the supervised short-circuit limitation algorithms.